\documentstyle[12pt,epsf]{article}

\newcommand \be  {\begin{equation}}
\newcommand \ee  {\end{equation}}
\newcommand \bea {\begin{eqnarray} \nonumber}
\newcommand \eea {\end{eqnarray}}

\newtheorem{theorem}{Theorem}

 \let\lb=\label
\def\pref#1{(\ref{#1})}
\def\II#1{{\cal I}^{(#1)}}

\let\a=\alpha \let\b=\beta  \let\g=\gamma    \let\d=\delta  
       
\let\m=\mu    \let\n=\nu           \let\r=\rho
\let\s=\sigma \let\t=\tau     \let\o=\omega
 \let\D=\Delta    \let\L=\Lambda
    \let\F=\Phi         \let\O=\Omega

\begin{document}

\title{Some Numerical Results on the Block Spin Transformation for
the $2D$ Ising Model at the Critical Point}

\author{
G. Benfatto\thanks{Dipartimento di Matematica, Universit\`a di Roma ``Tor
Vergata'', I-00133 Roma, Italy - Partially supported by MURST (fondi di
ricerca 40\%) and CNR-GNFM.}\\
{\footnotesize \tt benfatto@mat.utovrm.it}
\and
E. Marinari\thanks{Dipartimento di Fisica and INFN, Universit\`a di Roma ``Tor
Vergata'', I-00133 Roma, Italy, and NPAC, Syracuse University, Syracuse NY
13210, USA}\\ {\footnotesize \tt marinari@roma1.infn.it}
\and
E. Olivieri$^*$\thanks{Work partially supported by Grant SC1-CT91-0695 of
the Commission of European Communities.}\\
{\footnotesize \tt olivieri@mat.utovrm.it}
}

\date{April 1994}
\maketitle
\thispagestyle{empty}
\begin{abstract}

We study the block spin transformation for the 2D Ising model at the critical
temperature $T_c$. We consider the model with the constraint  that the total
spin in each block is zero. An old argument by Cassandro and Gallavotti allows
to show that the Gibbs potential for the transformed measure is well defined,
provided  that such model has a critical temperature $T'_c$ lower than $T_c$.
After describing a possible rigorous approach to the problem, we present
numerical evidence that indeed $T'_c<T_c$, and a study of the
Dobrushin-Shlosman uniqueness condition.

\end{abstract}

\vfill



\vfill

\newpage

\setcounter{page}{1}
\section{Introduction\protect\label{S_INT} }

In this note we discuss the block spin transformation for the two dimensional
Ising model at the critical point, trying to show that it is well defined, and
it gives rise to a Gibbs measure corresponding to a translationally invariant
finite norm potential. If one wants to define the renormalization group flow in
the space of Hamiltonians this is of course an essential point. Here we study
only the first step of this flow.
Similar results have been obtained by Kennedy \cite{Ke} for the majority rule
renormalization group transformation.

Our aim would be to use a rigorous approach to the subject but, as it will
appear clear in the sequel, this is presently a very difficult task. Hence we
will mainly analyze the problem from a numerical point of view.  However, as
we will explain, this will be partially achieved by combining theoretical
perfectly rigorous ideas with numerical tools in order to ``measure'' via a
computer some interesting theoretical quantities.

The problem of considering well defined renormalization group transformations
(RGT) and, in particular, the study of RGT in the framework of the modern
rigorous approach to statistical mechanics, has attracted the attention of
several authors.

Recent and less recent papers have been dedicated to this subject.  We want to
quote in particular, among the ``old'' papers, the one by Cassandro and
Gallavotti \cite{CG} (which, at our knowledge, is the first one treating
explicitly the above mentioned problem), the paper \cite{GP} by Griffiths and
Pearce and the one by Israel \cite{I}.  For the recent results we quote the
monumental paper \cite{EFS} by van Enter, Fernandez and Sokal, where the
problem
is discussed in a very clear and complete way.  This article-book is reach of
very interesting discussions and examples. It contains a self-contained
exposition of the general set-up and a very extensive and up-to-date
bibliography.  Thus we refer the interested reader to \cite{EFS} for a review
on
the subject and for reference to other recent papers.

It is worthwhile to remark that the majority of the examples considered in the
above papers, concerning the rigorous approach to RGT, deal with the region
far (and often very far) from the critical point.  The paper \cite{CG}
constitutes, in some sense, an exception and some crucial ideas developed in
the present paper go back, in fact, to \cite{CG}.

In the following section we shall present a general critical discussion about
the definition of RGT.  In section \ref{S_MOD} we shall give the definitions
concerning the models considered in this paper and the Monte Carlo procedure
used to study their equilibrium properties. Section \ref{S_INEQUA} will be
devoted to the Dobrushin-Shlosman uniqueness condition and to the exact
definition of a related numerical quantity, that we have studied
to analyze the RGT (see next section). Sections \ref{S_PT} and
\ref{S_BC} contain the numerical results and, finally, section \ref{S_CON} is
devoted to the conclusions.

\section{\protect\label{S_GD} A Critical Discussion About the RGT}

The main question arising when rigorously discussing the RGT can be
explained by considering, for example, the Ising model at magnetic field $h$
and
inverse temperature $\b$. Let
$$
  \nu\,=\,T_{(b)}\,\mu_{\beta ,h}
$$

\noindent be a measure arising from the application of a renormalization group
transformation $T_{(b)}$, defined ``on scale $b$'', to the Gibbs measure
$\mu_{\beta ,h}$ of the Ising model. Transformations of this kind are always
trivially defined in finite volume but, of course, we are interested in taking
the thermodynamic limit or, rather, in defining directly the transformation in
the infinite volume situation.

The main ``pathology'' that can take place, which is the main part of the
discussions contained in \cite{EFS}, is that $\n$ can be non-Gibbsian. This
means that the conditional probabilities of $\n$ can be incompatible with the
Gibbs prescription corresponding to any absolutely summable potential. This
non-Gibbsianness is detected via the violation of a necessary condition, namely
the property of {\it quasi-locality } for the conditional probabilities of
$\n$.

The Dobrushin--Lanford--Ruelle (DLR) theory of Gibbs measures is bas\-ed on the
conditional probabilities $\pi_{\Lambda}$ for the behavior of the system in a
finite box $\L\subset \subset {\bf Z^d}$, subject to a specific configuration
in
the complement of $\L$ (we use the notation $\L\subset \subset {\bf Z^d}$ to
denote a {\it finite} subset $\L$ of $ {\bf Z^d}$).  For simplicity let us only
consider Ising-like systems.  The configuration space of the system, in this
case, is $\Omega = \{-1,1\}^{{\bf Z^d}}$; we use $\Omega_{\Lambda} =
\{-1,1\}^{\Lambda}$ to denote the configuration space in $\Lambda
\subset{\bf Z^d}$.  According to \cite{EFS}, a probability measure whose
conditional probabilities $\{ \pi_{\Lambda} \}_{\Lambda \subset \subset {\bf
Z^d}}$ satisfy:

\be
  \lim _{\Lambda'\uparrow {\bf Z^d}}\ \
  \sup _{\omega_1,\omega_2 \in \Omega:
  (\omega_1)_{\Lambda'}=(\omega_2)_{\Lambda'}}
  |\pi_{\Lambda}f(\omega_1) - \pi_{\Lambda}f(\omega_2)|=0
  \lb{1.1}
\ee

\noindent (namely the conditional expectations in $\Lambda$ of any cylindrical
function $f$ corresponding to different boundary conditions
$\omega_1,\omega_2$,
coinciding in $\Lambda' \, \supset \, \Lambda$, tend to coincide as $\Lambda'$
tends to ${\bf Z^d}$ ), is called {\it quasilocal}. \pref{1.1} can be seen as a
continuity property in the conditioning, infinite volume, configuration $\o$.

Kozlov has shown in \cite{Ko} that a quasilocal probability measure on $\Omega$
which also satisfies a so called {\it non-nullity} condition, i.e. a sort of
absence of hard core exclusion, is {\it Gibbsian}, in the sense that its
conditional probabilities can be obtained, via the Gibbs prescription, from an
absolutely summable potential (see \cite{EFS} for more details).

It is useful to make at this point some remarks about this notion of
quasilocality.

\vspace{.3cm}
\noindent 1) \ Kozlov's theorem (i.e. the fact that nonnullness together with
quasilocality
imply Gibbsianness)  is proved in \cite{Ko} by using an approach which  can be
considered somehow artificial. Starting {\it only} from some nice continuity
properties of the conditional probabilities $\{\pi_{\Lambda}\}$, one gets a
series, representing the interaction of a point $x$ with the rest of the world,
which is, a priori, only {\it semi-convergent}. One can insist to extract  the
many body potentials from that series (for instance via the Moebius inversion
formula) pretending that they are absolutely summable. This can be achieved
only
by regrouping the terms in some suitable order. One can use, for instance, the
lexicographic order of the lattice.  The resummation will depend in this case
on
the location of $x$. The potential will be absolutely summable but, in
general, not translationally invariant. To get translational invariance one
needs some stronger properties on how weakly the conditional probabilities
depend on far apart configurations.

In some situations to compute the renormalized potentials one can use much
stronger methods, based on convergent cluster expansions; in this way a genuine
finite norm, translationally invariant, potential is produced in a very natural
way. Each {\it renormalized coupling constant} is expressed via a convergent
series (see, as an example, the paper \cite{C} by Cammarota).

\vspace{.3cm}
\noindent 2) \ The above notion of quasilocality of a measure $\m$ needs a
control {\it
uniform } in the conditioning configuration.  To prevent the existence of a
Gibbs potential it is sufficient that the condition is violated for
{\it only one} special configuration;
in this case even the somehow artificial quantity introduced in \cite{Ko}
cannot
be constructed.  However, for any infinite volume reasonable stochastic field
$\m$, a {\it single} infinite volume configuration $\o$ is of zero measure;
moreover it can even happen (see below) that the single configuration inducing
non-Gibbsianness, as a consequence of
non-quasi-locality, is very ``non-typical''
with respect to $\m$.  It is then natural and physically relevant to introduce
a
weaker notion of quasilocality, for instance by requiring the validity of a
condition like \pref{1.1} only for $\m$-almost all $\o$'s.

A precise definition in this sense has been recently introduced by Fernandez
and Pfister (see \cite{FP}), but this does not prevent the construction of
pathological examples.  In fact in \cite{FP} the authors show that, for some
interesting examples, a very strong notion of non-quasilocality holds, in the
sense that \pref{1.1} fails actually {\it for $\m$-almost all }
configurations $\o$.

This happens, for instance, for the example of non-Gibbsianness given by
Schonmann in \cite{S}.  This example consists simply in considering the {\it
relativization} $\n$ of the measure $\m^+$ for the Ising model in two
dimensions to the line $l = \{x\equiv (x_1,x_2) \; \in \; {\bf Z^2} \; :\; x_2
= 0\}$ (isomorphic to the one dimensional lattice $ {\bf Z^1}$).  Here we
are at large inverse temperature $\b$ and zero magnetic field; $\m^+$ is one
of the two extremal Gibbs measures, the one obtained via a thermodynamic limit
with + boundary conditions; finally, by relativization of $\m^+$ to $l$ we
simply mean the projection on the $\s$-algebra generated by the spins in $l$
or, simply, the (marginal) distribution of the spins in $l$ deduced by $\m^+$
by integrating out all the spins in $ {\bf Z^2 }\setminus l$.

\vspace{.3cm}
\noindent 3) \ There are many cases in which a stochastic field $\m$ shows
up the pathology of
non-Gibbsianness, like in the Schonmann example; however, at the same time, the
measure $\m$ can be, from many respects, very well behaved. For instance the
one
dimensional measure $\n$ of the Schonmann example has exponentially decaying
truncated correlations. Moreover, as it has been recently shown by Lorinczi and
Vande Velde (see \cite{LvV}), this even very strong non-Gibbsianness is, in a
sense, an {\it unstable} property.

Let us reanalyze the Schonmann example. Suppose that,  instead of considering
the one dimensional sublattice $l$, one considers a sublattice $l_b$ of
sufficiently large spacing $b$. Namely starting from $\m^+$, one integrates out
all the spins outside the set $l_b \equiv \{ x \equiv (x_1,x_2) \; \in \; {\bf
Z^2} \; :\; x_2 = 0, x_1 = nb, n \; \in {\bf Z^1}\}$, obtaining the relativized
measure $\n_b$ on $l_b$.
$\n_b$ can also be seen as obtained via a {\it decimation
procedure } on scale $b$ from $\n$.
In general, given a measure $\m$ on $\Omega = \{-1,1\}^{{\bf Z^d}}$ and an
integer $b\geq 2$, the decimation transformation $T_b$ acts on $\m$ so that
$$ \n = T_b\m $$
is simply the relativization of $\m$ to the sublattice ${\bf Z^d_b}$ of
${\bf Z^d}$ with spacing $b$, that is ${\bf Z^d_b}= \{ x \; \in \;{\bf Z^d} \;
:\; x= b y ,\; y\; \in \; {\bf Z^d} \} $.
Lorinczi and Vande Velde show that $\n_b$ is Gibbsian in the strong
sense, that the renormalized potential can be computed via a cluster expansion
and it is absolutely convergent.

Another interesting example given in \cite{EFS} concerns the decimation
transformation applied to the unique Gibbs measure $\m_{\b,h}$ for the Ising
model at large $\b$ and $h \ne 0,$ say $ h >0$.  They show that, $\forall \; b$
and for suitable $\b $ and $h$, the renormalized measure $ \n = T_b\m_{\b,h}$,
arising from a decimation transformation with spacing $b$, is not consistent
with any quasi-local specification.  In particular it is not the Gibbs measure
for any uniformly convergent interaction. As it is noticed in \cite{EFS} the
non
existence of the renormalized interaction is a consequence of the presence of a
first order phase transition for the original model in ${\bf Z^d}\setminus {\bf
Z^d_b}$ for particular values of $(\omega_x)_{x\in {\bf Z^d_b}}$ and suitable
values of $h$ and $\beta$. One example is the case where $\omega_x = -1$
$\forall
x $, $h$ uniform and positive, exponentially in $\beta$ near to the value
$h^*(b)$ which is needed to compensate, in ${\bf Z^d}\setminus {\bf Z^d_b}$,
the
effect of the $-1$'s in ${\bf Z^d_b}$ and to give rise to a degeneracy in the
ground state in ${\bf Z^d}\setminus {\bf Z^d_b}$ (the highly nontrivial part in
the proof, given in \cite{EFS}, of the existence of the pathology consists in
showing, via the Pirogov-Sinai theory, the persistence of the phase transition
at positive temperatures).

On the other hand from the above analysis it is clear
that this pathology comes
from the fact that, on a too short spatial scale $b$ (with respect to the
thermodynamic parameters and mainly to the magnetic field $h$), the system is
reminiscent of the existence of a phase transition for $h = 0$. It seems
reasonable that this pathology could be eliminated provided one uses a RG
transformation defined on a proper scale depending on the thermodynamic
parameters. In \cite{MO4} Martinelli and Olivieri have shown, exactly for the
above example of Ising model for which in \cite{EFS} the pathology is found,
that, with the same values of $\b$ and $h$, provided one chooses a
sufficiently
large spacing $b' >b$, the resulting measure $T_{b'}\m_{\b,h}$ is Gibbsian in
the strong sense and that the renormalized potential, which is absolutely
summable, can be computed via a convergent cluster expansion.  In particular,
taking $b' = b^n $, with $n$ sufficiently large, one shows that, iterating $n$
times the transformation $T_b$, one goes back to Gibbsian measures; moreover it
has also been shown in \cite{MO4} that $T_{b^n}\m_{\b,h}$ converges, as $n$
tends to $\infty$, to a trivial fixed point.

\vspace{.3cm}

Let us now describe an example of pathology discussed in \cite{EFS} which is
particularly relevant in the context of the present paper. It refers to the
block averaging transformation (sometimes called Kadanoff transformation).
Suppose to partition ${\bf Z^2}$ into square blocks $B_i$ of side $2$ (each
block containing 4 sites). The block averaging transformation $T^B_{(2)}$
consists, in this case, in the following transformation applied to the Gibbs
measure $\m_{\b,h}$ for the Ising model at inverse temperature $\b$ and
magnetic
field $h$; the new measure is obtained, starting from the original spin
variables
$\s_x$, by assigning to any block $B_i$ an integer value $m_i$ and by computing
the probability, with respect to the original Gibbs measure $\m_{\b,h}$, of the
event $ \sum _{x\in B_i} \s _x = m_i$.

One obtains, starting from $\m_{\b,h} (\{\s_x\})$:

$$
  \n (\{m_i\}) = T^B_{(2)}\m_{\b,h}
$$

\noindent The original system of $\s_x$ variables distributed according to
$\m_{\b,h}$ is called {\it object system}, whereas the new variables $m_i$
distributed according to $\n$ constitute the {\it image system}.

The pathology in the block averaging transformation for the Ising model at
large
inverse temperature $\b$ and arbitrary magnetic field $h$ is a consequence of
the existence of a phase transition for the object system for particular values
of the image variable $m_i$.  The authors of \cite{EFS} show that for the
configuration with $m_i=0$, $\forall \; i$, the corresponding object system, a
constrained Ising model, exhibits a phase transition with long range order. As
a
consequence of this fact they are able to show the violation of the
quasi-locality condition.

Of course, since the local magnetizations $m_i$ in the blocks $B_i$ are fixed
and all equal to zero, the value of $h$ is totally irrelevant.  On the other
hand if $h$ is very large and, say, positive, the object system without any
constraint is almost Bernoulli with a high probability to have an individual
spin equal to $+1$ and the configuration with $m_i=0$, $\forall \; i$, is
expected to be very unlikely and, in a sense, irrelevant. Probably the weaker
condition of almost sure quasi-locality introduced in \cite{FP} is satisfied in
that situation. Moreover, even though $\n$ is not Gibbsian, it could have nice
properties and its non-Gibbsianness could be unstable with respect to small
changes. The situation could be similar to the previous mentioned phenomenon
discovered by Lorinczi and Vande Velde for the decimation applied to the
measure
appearing in the Schonmann example.

It is interesting now to discuss in some detail one of the main ideas contained
in the paper \cite{CG} which is, in a sense, at the basis of the present note.
We will consider the  block averaging transformation we have just discussed,
and
apply it to the Ising model  Gibbs measure $\m_{\b,h}$. The authors of
\cite{CG}
are concerned, in particular, with the most interesting example where $h=0$ and
$\b =\b_c$, i.e. the system is at the critical point. They show that, at
least formally, it is possible to compute the renormalized potential and to
show
that it is absolutely summable, provided the constrained Ising model with all
the $m_i=0$, $\forall \; i$, is above its critical temperature.

To be more precise, let $H^{(r)} (\{m_i\})$ be the renormalized Hamiltonian
corresponding to the renormalized measure $\n (\{m_i\}) = T^B_{(2)}\m_{\b,h}$;
suppose to extract from $H^{(r)} (\{m_i\})$ all the many-body potentials $\F_A
(\{m_i\}_{i\in A})$ for any finite set $A$ of blocks $B_i$. The authors of
\cite{CG} show that all the $\F_A $'s can be expressed as thermal averages of
suitable local observables with respect to the Gibbs measure corresponding to
an auxiliary intermediate Hamiltonian, that we call $H^{(6)}(S)$. $H^{(6)}(S)$
corresponds to the constrained Ising model with all the $m_i$'s set equal to
zero. The new (intermediate) local variables $S_i$, defined for any block
$B_i$,
take values in the finite space, containing six states, corresponding to the
six
spin-$\s$ configurations in $B_i$ such that $ \sum _{x\in B_i}\s_x =0$.

The starting point of the present note is to try to rigorously show, in a
strong
sense, that the auxiliary model with Hamiltonian $H^{(6)}(S)$ does not undergo
a  phase transition at $\b = \b_c$. A priori there are no reasons, as it will
be clear from the discussion in the following sections, for the  critical
temperature to decrease after the introduction of additional constraints to a
spin model. We will show for example that, as a consequence of a remark due to
Kasteleyn, a particular constrained model obtained from the Ising zero field
model has  {\it exactly} the same critical temperature as the original Ising
model~!

In particular, to detect the absence of phase
transition, we will use the idea of
exploiting some {\it finite size condition}, which goes back to Dobrushin and
Shlosman (see \cite{DS1,DS2,DS3}).  The basic point is that if one is able to
verify a condition involving mixing properties of (finite volume) Gibbs
measures
in a suitable set of finite regions, then one can deduce nice properties
(typical of the one phase region) for the {\it infinite volume} system. That
can
be done, for example, by using a computer. One can show for example  uniqueness
of the infinite volume Gibbs measure, analyticity of the infinite volume
thermodynamic and correlation functions and exponential decay of truncated
correlations.   In \cite{DS1,DS2,DS3} the authors avoid the use
of cluster expansion; in \cite{DS2,DS3} they use conditions referring to
arbitrary shapes.

In \cite{DS1} the authors introduce a somehow weak condition implying only
uniqueness of the infinite volume Gibbs state and some decay properties of the
infinite volume truncated correlations.This condition refers to a region
$V\subset \subset {\bf Z^d}$ and is usually called $DSU(V)$ from
Dobrushin-Shlosman uniqueness condition (see \pref{2.6}, \pref{2.7} below). In
\cite{DS2,DS3} they treat the so called {\it completely analytical}
interactions, proving, on the basis of a stronger condition, much stronger
results, in particular uniform analyticity and exponential decay of truncated
correlations {\it for any} finite or infinite volume $\L$ with constants
uniform
in $\L$.

In \cite{O,OP} Olivieri and Picco consider similar finite size conditions but
only for sufficiently regular regions and get similar results of strong type
(like Dobrushin-Shlosman complete analyticity) by using a block decimation
procedure and the theory of the cluster expansion. In a series of papers
(\cite{MO1,MO2,MO3,MO4}) Martinelli and Olivieri developed a critical analysis
of the known finite size conditions getting new results both for the
equilibrium
(Gibbs state) and for the non-equilibrium (Glauber dynamics) situation. The
theory developed in \cite{MO1,MO2,MO3} allows, contrary to the
Dobrushin-Shlosman analysis, to treat, for quite general lattice systems,
almost
the whole one-phase region (see, in particular, \cite{MO2} for more details).

In a recent paper \cite{MOS} Martinelli, Olivieri and Schonmann showed that,
{\it in two dimensions}, two finite volume mixing conditions of a priori
different strength called, respectively, Weak Mixing (WM) and Strong Mixing
(SM)
conditions, are in fact equivalent for sufficiently regular domains $\L$ (see
\cite{MOS} for more details). In \cite{DS1} the authors show that if there
exists a a region $V\subset \subset {\bf Z^d}$ such that their finite size
condition $DSU(V)$ is satisfied, then weak mixing holds for any finite or
infinite $\L \subset {\bf Z^d}$. Then, combining the results in \cite{DS1} with
the ones in \cite{MOS} one gets that in two dimensions, if there exists a
finite
region $V\subset \subset {\bf Z^2}$ such that $DSU(V)$ is satisfied for the
constrained Ising system with Hamiltonian $H^{(6)}(S)$, then, for a large class
of regular domains, including for instance any cube,the strong mixing condition
is satisfied for this constrained system.

 From strong mixing, using the results obtained in \cite{O,OP}, one can easily
make completely rigorous the above mentioned argument introduced by Cassandro
and Gallavotti, and compute the renormalized potentials {\it as convergent
series} via the cluster expansions.  In this way after proving $DSU(V)$ the
Gibbsianness of the renormalized measure would be proven in the strongest
possible sense.

\section{\protect\label{S_MOD}
Definition of the Models and of the Heat Bath Dynamics}

In the following we will give precise definitions about some models that we
are going to discuss: the usual 2D Ising model and some ``restricted'' models
obtained from the Ising model by imposing some ``extensive'' restrictions.

Suppose to partition ${\bf  Z^2}$ into $2 \times 2$  squared blocks $B_i$, each
containing $4$ sites. Each block $B_i$ can be characterized by the coordinates
of its lower left-hand site $y_i$; namely $B_i \equiv \; B_{y_i}$,  where $y_i
=
2 x_i, \; x_i \; \in \; {\bf  Z^2}$ and for $x \equiv (x^{(1)},x^{(2)}) \in
{\bf  Z^2}$ :

$$
  B_{2x}\; = \; \{ z \; \in \; {\bf  Z^2} : 2x^{(j)} \leq z^{(j)} <
  2( x^{(j)}+1) , \;\; j =1,2\}
$$

\noindent The formal Hamiltonian associated to the usual Ising model in zero
magnetic field is given by:

\be
  H^{Ising} \equiv - \sum_{<x,y>} \sigma_x \sigma_y\ , \lb{1.5a}
\ee

\noindent where the sum runs over the pairs of nearest neighbors sites in ${\bf
Z^2}$, and $ \sigma_x \in \; \{ -1, +1\}$.

In the following we will consider a system enclosed in a finite squared region
$\Lambda$ with various boundary conditions; if not explicitly specified, it
will be understood that the boundary conditions are periodic.

In the original Ising model there are, in each block $B_i$, 16 allowed
configurations.  Instead of the original $\s_i$ variables to describe such
configuration we can as well use  the block variables, say $S_i \; \in \; \{ 1,
\dots, 16\}$ . In each block there will be a self-interaction and the mutual
interaction between blocks deriving from \pref{1.5a} is again of
nearest-neighbor type. The Hamiltonian of the Ising model, expressed in terms
of $S_i$'s block variables, will be denoted by $H^{(16)} (S) \; \equiv \;
H^{Ising} (\sigma)$.

We will consider a modified model in which in each block $B_i$  the sum
$ m_i = \sum _{x\in B_i} \sigma_x$ is constrained to be zero.
Now the block variables $S_i$ will assume only $6$ different values,
corresponding to the following six block configurations:

\be
  \protect\label{E_BL6}
  \left [ \begin{array}{cc} + & +\\ - & - \end{array} \right ]\ ,
  \left [ \begin{array}{cc} - & +\\ - & + \end{array} \right ]\ ,
  \left [ \begin{array}{cc} - & -\\ + & + \end{array} \right ]\ ,
  \left [ \begin{array}{cc} + & -\\ + & - \end{array} \right ]\ ,
  \left [ \begin{array}{cc} + & -\\ - & + \end{array} \right ]\ ,
  \left [ \begin{array}{cc} - & +\\ + & - \end{array} \right ]\
\ee

\noindent
The corresponding Hamiltonian will be denoted by $ H^{(6)} (S)$.

It is easy to convince oneself that the model with Hamiltonian $H^{(6)}(S)$
has four periodic ground states (see \cite{EFS}). The first one is given by:

\be
  \protect\label{E_BL2} \nonumber
  \begin{array}{cccccccc}
+ & + & + & + & + & + & + & + \\
- & - & - & - & - & - & - & - \\
- & - & - & - & - & - & - & - \\
+ & + & + & + & + & + & + & +
\end{array}
\ee

\noindent The second one is obtained from the first one by interchanging the
$+$ with the $-$; the last two ground states are obtained from the first two
by interchanging the rows with the columns.
Note that the last two block configurations in \pref{E_BL6} are quite
different from the first four; in fact they are the only ones that carry a
non-zero internal energy and they are absent from the $T=0$
ground state structure. We call them {\bf turnons}, since, as we
will see, they play an important role by allowing the layered
ground states to break and mix, destroying long range order. They are

\be
  \nonumber
  \left [ \begin{array}{cc} + & -\\ - & + \end{array} \right ]\ ,
  \left [ \begin{array}{cc} - & +\\ + & - \end{array} \right ]\ .
\ee

\noindent By further restricting the allowed block configurations, so to forbid
the presence of turnons, we define a new model whose Hamiltonian will be
denoted
by $H^{(4)}(S)$.

In the following we shall denote the three models introduced before by
$\II{n}$,
with $n=4,6,16$. In section \ref{S_EQUIVA} we will show that $\II4$ with
periodic boundary conditions is exactly equivalent to two uncoupled Ising
models
(in a smaller volume); hence its critical temperature is exactly the same as in
$\II{16}$.

We have studied the $3$ models $\II{n}$ by a Monte Carlo procedure, based on
a suitable {\em Heat Bath dynamics}, whose invariant distribution is the
finite volume Gibbs measure. This dynamics has been used to compute
mean values (with respect to the Gibbs measure) of some relevant observables
as time averages; the mean value will be denoted by $\langle \cdot \rangle$ in
the following.

We have built a discrete time Heat Bath dynamics based on locally
equilibrating the $S_i$ block variables.  We suppose that $\Lambda$ is a cube
of even side size, so that it can be exactly partitioned into $N(\L)$ $B_i$
blocks, ordered in a lexicographic way, and we define $ \Omega ^{(n)}
_{\Lambda} = \{ 1, \dots , n\}^{N(\Lambda)}$.  The dynamics in a finite volume
$\Lambda$ is given by a Markov chain defined below.

We perform a complete update of all the $N(\L)$ block variables by
successively updating each one of them.  For any $i=1,\ldots,N(\L)$, we choose
at random the new block configuration $S'_i$ in $B_i$, given the
configuration $\{S_k\}_{k=1}^{N(\L)}$ in $\L$, according to
the equilibrium Gibbs measure in $B_i$ with $S|_{B_i^c}$ boundary conditions
($B_i^c = \Lambda \setminus B_i$). The related transition probability in the
$\II{n}$ model is then given by:

$$
  P^{(n)}(S\to S')=
  { \exp [- \beta H^{(n)}(S'_i|S_{B_i^c})] \over
  \sum_{S''_i=1}^n \exp [- \beta H^{(n)}(S''_i|S_{B_i^c})] } \ .
$$

\noindent This is the most efficient method as far as local updates of block
variables are concerned (since locally we bring at equilibrium the basic block)
and is implemented by means of a simple look-up table.  It is straightforward
to
build up the dynamics so that one can move from one model to another simply
changing the number of allowed states (the $6$ blocks of (\ref{E_BL6}) are
stored in the first $6$ positions of the tables, with the two turnons in
position $5$ and $6$).

In order to characterize the critical point of the system we have
computed two different quantities. First of all we have considered the
specific heat $C_\L$ as defined from the equilibrium energy fluctuations:

\be
  \protect\label{E_CV}
  C_\L \equiv |\L|^{-1} \beta^2
  (\langle H^2 \rangle - \langle H \rangle^2)\ .
\ee

\noindent We have also considered a correlation length $\xi$  defined by
measuring zero-momentum correlation functions. We define the sums over planes

\be
  \tilde{m}(t) \equiv \sum_{s=1}^L \s_{(t,s)}\ ,
\ee

\noindent where $x= (x^{(1)},x^{(2)})\equiv (t,s)$, and the correlations

\be
  G(t) \equiv \frac{1}{|L-t|}
  \sum_{t_0=1}^{L-t} \langle \tilde{m}(t_0) \tilde{m}(t_0+t) \rangle\ .
\ee

\noindent A $t$-dependent correlation length (which we will plot
for $t=5$, where we get a fair estimate for its limit as
$t\to\infty$) can be defined by

\be
  \xi(t) \equiv ( \log \frac{G(t)}{G(t+1)} )^{-1}\ . \lb{corr}
\ee

\noindent For the constrained models, which have a pathological behavior
at odd separations, we have found practical to define the correlation
length by a distance $2$  ratio

\be
  \xi_{(2)}(t) = ( \frac{1}{2} \log \frac{G(t)}{G(t+2)} )^{-1}\ . \lb{corr2}
\ee

\noindent Note that in the $\II{16}$ model $\xi(t)$
coincides with $\xi_{(2)}(t)$ for $t\to\infty$.

\subsection{ \protect\label{S_EQUIVA}
$4$ State Restriction and Full Ising Model: Proof of the
Equivalence}

Let us denote ${\cal B}_L$ the family of $2\times 2$ blocks partitioning the
cube of (even) side size $L$ and $\{S_\alpha\}_{\alpha\in {\cal B}_L}$ the
generic configuration of the $\II4$ model. We have:

\be
  H^{(4)}(S) = \sum_{<\alpha,\beta>} w(S_\alpha,S_\beta)
\ee

\noindent where $<\alpha,\beta>$ denotes a couple of
nearest neighbor blocks and
$w(s_\alpha,s_\beta)$ is the interaction energy between the two blocks (the
self interaction vanishes).

We want to show that it is possible to associate to each block $\alpha$ an
invertible map $\phi_\alpha: S_\alpha \to (\rho_\alpha,\tau_\alpha)\in
\{-1,+1\}^2$, so that, for any couple $<\alpha,\beta>$ and any choice of
$S_\alpha,S_\beta$:

\be
  \protect\label{EINT} w(S_\alpha,S_\beta)=\rho_\alpha \rho_\beta +
  \tau_\alpha \tau_\beta
\ee

\noindent We first observe that all the four block
configurations are of the form

\be
     \left[ \begin{array}{cc} \sigma_1 & \sigma_2\\
     -\sigma_2 & -\sigma_1 \end{array} \right]
\ee

\noindent Hence there is a simple way to define $\phi_\alpha$ on the blocks
contained in a $4\times 4$ square, so that (\ref{EINT}) is verified at least
for
the blocks in the square. This definition can be visualized in the following
picture:

\be \protect\label{SQ}
\begin{array}{cc}
     \left[ \begin{array}{cc} -\tau_2 & -\r_2\\
     \ \r_2 & \ \tau_2 \end{array} \right] &
     \left[ \begin{array}{cc} -\r_3 & -\tau_3\\
     \ \tau_3 & \ \r_3 \end{array} \right] \\
              &   \\
     \left[ \begin{array}{cc} \ \r_1 & \ \tau_1\\
     -\tau_1 & -\r_1 \end{array} \right] &
     \left[ \begin{array}{cc} \ \tau_4 & \ \r_4\\
     -\r_4 & -\tau_4 \end{array} \right]
\end{array}
\ee

\noindent Moreover, if $L$ is a multiple of $4$, there is a partition of the
lattice into $4\times 4$ squares and it is immediate to check that
(\ref{EINT}) is
satisfied for all couples $<\alpha,\beta>$, if $\phi_\alpha$ is defined in
each $4\times 4$ square as in (\ref{SQ}), by periodic extension.

To complete the proof it is sufficient to observe that (\ref{EINT}) implies
the following identity for the partition functions:

\be
  Z_{\II4,L}(\beta) = Z_{\II{16},L/2}(\beta)^2
\ee

The model $\II4$ has four ground states, exactly coinciding
with the ones of $\II6$ (the turnons in this case are absent at $T=0$).
These four states correspond, via the map $\phi_\a$, to the four ground
states of the two independent Ising models (all $+1$ or all $-1$ for each one
of the two Ising systems).

\section{ \protect\label{S_INEQUA}
The \ \ Dobrushin-Shlosman \ \ Uniqueness \ \ Condition}

Let us define the {\it variation distance} between two probability
measures $\m_1$ and $\m_2$ on a finite set
$Y$\footnote{a much more general framework can also be considered} as:

\be
  \hbox{Var} (\m_1, \m_2) \,=\,
  {1\over 2} \sum_{y\in Y}\vert \m_1(y) - \m_2(y)
  \vert \,=\,
  \sup_{X \subset  Y}\vert \m_1(X) - \m_2(X)\vert  \lb{1.6}
\ee

\noindent Given a
metric $\r(\cdot, \cdot)$ on $Y$ the {\it Kantorovich - Rubinstein - Ornstein -
Vasserstein distance with respect to $\r$} between two probability measures
$\mu_1$, $\mu_2$ on $Y$, that we denote by $V_\r(\mu_1, \mu_2)$, is defined as

\be
  V_\r(\mu_1, \mu_2) \,=\, \inf_{\mu\in K(\mu_1,
  \mu_2)}\sum_{y, y'\in Y} \r(y, y')\mu(y , y') \lb{1. 7}
\ee

\noindent where $K(\mu_1, \mu_2)$ is the set of joint representations of
$\mu_1$ and $\mu_2$, namely the set of measures on the cartesian product $Y
\times Y$ whose marginals with respect to the factors are, respectively,
given by $\mu_1$ and $\mu_2$.  This means that, $\forall B \subset Y$:

$$\mu(B\times Y) \,=\, \sum_{y\in B\, y'\in Y}\mu(y , y')\,=\, \mu_1(B)\ ,$$
$$\mu(Y\times B) \,=\, \sum_{y\in Y\, y'\in B}\mu(y , y')\,=\, \mu_2(B)\ .$$

\noindent For the particular case

\be \r(y , y') = \left\{ \begin{array}{ll}
1 & \quad \hbox{iff }y\ne y'\\ 0 & \quad \hbox{ otherwise} \end{array} \right.
\lb{1.8} \ee

\noindent it is possible to show that $V_\r(\m_1,\m_2)$ coincides
with the variation distance $ \hbox{Var}(\m_1,\m_2)$.
\bigskip

A result by Dobrushin and Shlosman \cite{DS1} concerning the uniqueness of the
infinite volume Gibbs measures generalizes previous results by Dobrushin based
on a ``one point condition'' on Gibbs conditional distributions (see
\cite{D2}).

Let us consider a spin system on ${\bf Z^d}$ with single spin space $\cal S$
and
finite range interaction. We generalize in an obvious way the notation
introduced in section \ref{S_GD}. Given a metric $\r$ on $\cal S$, we associate
to it a metric $\r_\L$ on $\O_\L\equiv {\cal S}^\L$, for any $\L \subset
\subset
{\bf Z^d}$, by defining:
$$\r_\L(S_\L,S'_\L) = \sum_{x\in\L} \r(S_x,S'_x)$$
We
say that condition $DSU_{\r} (\L, \delta)$ is satisfied if there exist a
finite set $\L \subset \subset {\bf Z^d}$ and a $\delta > 0$ such that
the following is true: for any $y \in \partial ^+ \L$ (the set of points
outside $\L$ whose spins interact with the spins inside $\L$) there is a
positive number $\a_y$ such that, for any couple of boundary conditions $\tau ,
\tau' \in \Omega^c_{\L}$ with $\tau'_x = \tau_x$, $\forall x \neq y$:

\be
  V_{\r_\L} (\mu^\tau_{\L},
  \mu^{\tau'}_{\L})\ \leq\ \alpha_y\r(\tau_y, \tau'_y)\ ,
  \lb{2.6}
\ee

\noindent and

\be
  \sum_{y\in\partial^+_r\L}\alpha_y
  \ \leq \ \delta\vert \L\vert\ .\lb{2.7}
\ee

\noindent where $\m_\L^\t$ is the Gibbs
measure in $\L$ with boundary conditions $\t$ outside $\L$.
We say that $DSU (\L, \delta)$ is satisfied if \pref{2.6} and
\pref{2.7} hold with $\r$ given by \pref{1.8}.

\begin{theorem} \lb{th1}
\quad(Dobrushin - Shlosman \cite{DS1})\quad
Let $DSU_{\r} (\L, \delta)$ be satisfied for some $\r$,
$\L$ and {\bf $\delta < 1$}; then $\exists\ \ C > 0$, $\gamma > 0$ such
that condition $WM_\r(\L, C, \gamma)$ holds {\it for every
$\Lambda $}.
\end{theorem}

\noindent By $WM_\r(\L,C,\g)$ we mean a particular mixing property of
$\m_\L^\t$, saying
that the influence at $x\in\L$ of a change in the conditioning spins $\t$
decays
as

$$
  C e^{-\g \hbox{dist} (x,\partial\L)}\ .
$$

\noindent See \cite{MO2} for a precise definition.

Theorem \ref{th1} implies, in particular, the uniqueness of infinite volume
Gibbs measures.  Then \pref{2.6}, \pref{2.7} provide an example of finite size
condition: one supposes that some properties of a {\it finite volume} Gibbs
measure are true and then deduces properties of {\it infinite volume}
distributions.

We observe now that

$$
  V_{\r_\L} (\mu^\tau_{\L}, \mu^{\tau'}_{\L}) \ge
  \sum_{x\in\L} V_\r(\m_x^\t, m_x^{\t'})\ ,
$$

\noindent where $\m_x^\t$ is the probability distribution of the spin $s_x$
with
respect to the measure $\m_{\L}^\t$. Hence, if $\r$ is given by \pref{1.8},
then

$$
  V_{\r_\L} (\mu^\tau_{\L}, \mu^{\tau'}_{\L}) \ge
  \sum_{x\in\L} \hbox{Var}(\m_x^\t, \m_x^{\t'})\ .
$$

\noindent It follows that, for given $\L$,
$DSU(\L,\d)$ is certainly \underline{not} satisfied for any $\d<1$, provided
that:

$$
  \sup_{\t,\t':\t_x=\t'_x,\forall x\not= y} \sum_{x\in\L}\hbox{Var}(\m_x^\t,
  \m_x^{\t'}) \ge \frac{|\L|}{|\partial^+\L|} \quad,\quad \forall y\in
  \partial^+\L
$$

\noindent This observation will be used in our numerical calculations on model
$\II6$ in the following way. We try to find a ``good'' lower bound for the
quantity

$$
  \sup_{y\in \partial^+\L_L} \sup_{\t,\t':\t_x=\t'_x,\forall x\not= y}
  \sum_{x\in\L_L} \hbox{Var}(\m_x^\t, \m_x^{\t'})
$$

\noindent by calculating numerically $\sum_{x\in\L_L} \hbox{Var}(\m_x^\t,
\m_x^{\t'})$ for a ``large'' number of choi\-ces of $y,\t,\t'$ and by
considering the maximum among these numbers, which we call $\D_L$.
Here $\L_L$ denotes a cube of side size $L$ in ${\bf Z^2}$ and each point of
$\L_L$ represents a $2\times 2$ block; hence $\L_L$ corresponds to a cube of
side size $2L$ in the original lattice.

We consider a ``reasonable'' indication that $DSU(\L,\d)$ is satisfied for
$\L$ large enough, if $\D_L |\partial^+\L_L|/|\L_L|$ is decreasing in $L$ and
there exists an $L_c$ such that

\be
  {\cal M}_{L} \equiv \D_{L/2} \frac{|\partial^+\L_{L/2}|}{|\L_{L/2}|} =
  \D_{L/2} \frac{8}{L} < 1 \quad,\quad L>L_c \ .\lb{DS}
\ee

\section{\protect\label{S_PT}
Numerical Results: the Phase Transition}

We will present here a first set of numerical results, which give
numerical evidence that the $\II6$ model undergoes a phase transition at
a temperature

$$
  T_c^{(\II6)} < T_c^{(\II{16})}\ .
$$

\noindent The correlation length of the constrained model at the critical point
of the full Ising model is a finite, reasonably small number, which we
estimate.

This numerical evidence is not meant to constitute a large scale simulation.
We do not try here to estimate critical exponents or to determine with high
precision the position of critical points (for large scale simulations of the
$2D$ Ising model see for example \cite{HEEBUR} and references therein).  The
main point here is to show in a non-ambiguous manner that the two critical
temperatures are different and that the correlation length of the constrained
model at the critical point of the Ising model is finite

Our results have been obtained for cubes containing $400^2$ lattice points
(i.e. the region $\L_L$ with $L=200$ according to the notation of section
\ref{S_INEQUA}).  In the case of model $\II6$ we have performed $2 10^5$ full
sweeps (that is full update of all lattice sites) of our Heat Bath block
algorithm per each value of the inverse temperature $\beta$.  In the case of
$\II4$ and $\II{16}$ models we have used $10^5$ sweeps per each value of
$\beta$.  We have simulated smaller volumes to check that everything was well
compatible with the expected finite size behavior (but we will not report in
detail about these data).  All the runs discussed in this Section use periodic
boundary conditions.

Let us start by discussing our simulations for the full $2D$ Ising
model. As we said, very large scale simulations exist ($\cite{HEEBUR}$)
and the results we are presenting here are just meant to set the frame
for showing numerically the different behavior of the two relevant
models. So we have simulated the full $2D$ Ising model in the same
conditions that we have used to study the constrained $\II6$ model.

\begin{figure}
\epsfxsize=400pt\epsffile{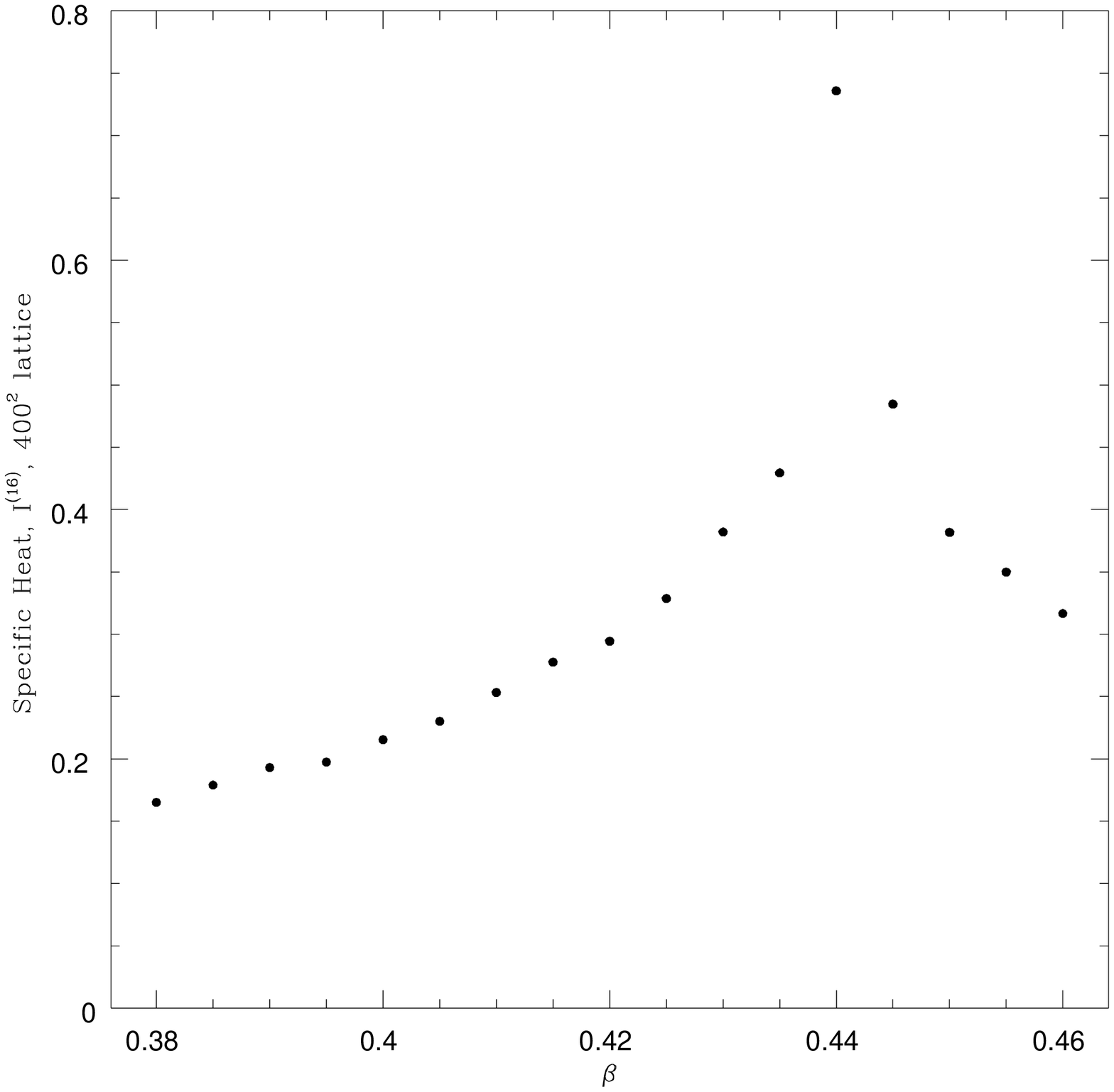}
  \caption[a]{\protect\label{F_CV16I}
    The specific heat (as computed from energy fluctuations), as a
    function of $\beta$ for the Ising model.
  }
\end{figure}

In fig.~\ref{F_CV16I} we show the specific heat of the model.  The point
closer to criticality is the one at $\beta=.4400$ (as is fortunately obvious
from the picture).  Here, as well as in the following, the point size is
of the order of magnitude of the statistical error, except for the point
closer to criticality. The mild (logarithmic) divergence of the specific heat
is at an inverse
critical temperature which we can estimate from our data to be at $\b_c = .440
\pm .001$, in agreement with the known exact value.

\begin{figure}
\epsfxsize=400pt\epsffile{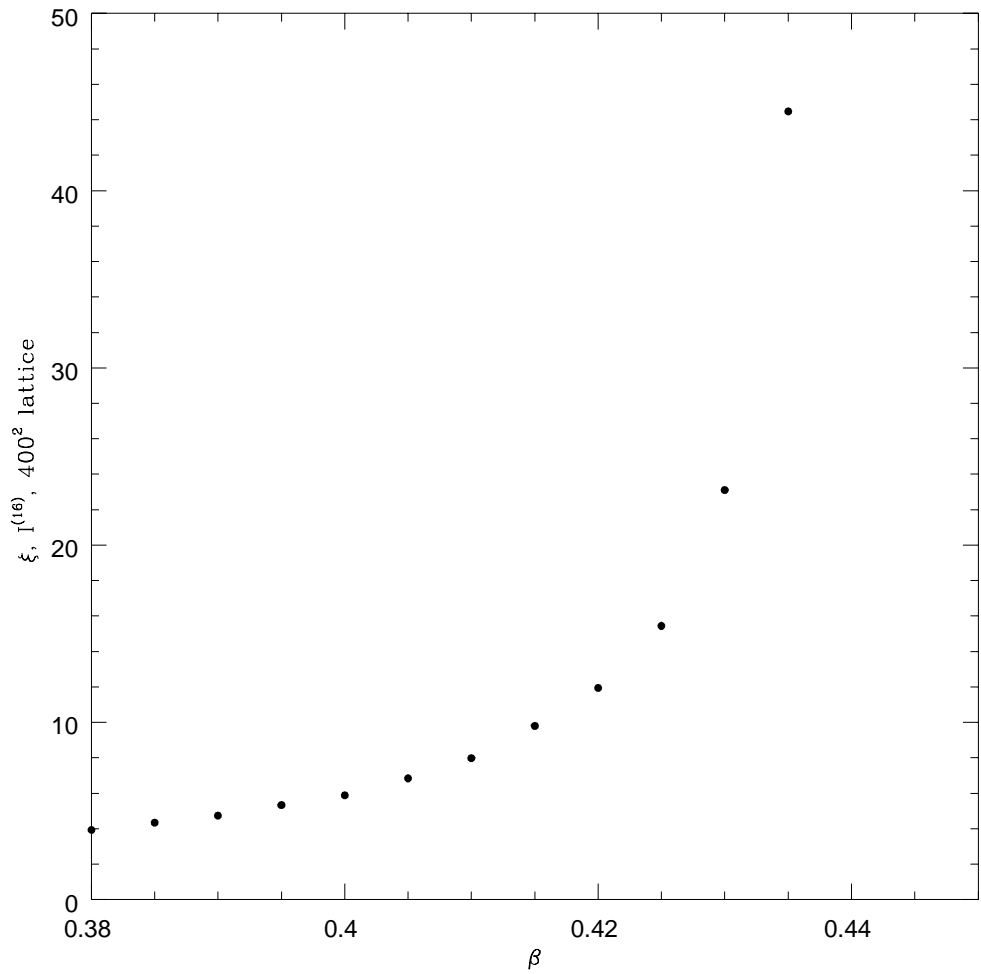}
  \caption[a]{\protect\label{F_CSI16I}
    The correlation length $\xi$ as a
    function of $\beta$, for the Ising model.
  }
\end{figure}

In fig.~\ref{F_CSI16I} we show the correlation length $\xi \equiv \xi(5)$ (see
\pref{corr}), which diverges at $T_c$ with a critical exponent $\nu=1$.  We
plot $\xi$ for $\xi \ll L$.

\begin{figure}
\epsfxsize=400pt\epsffile{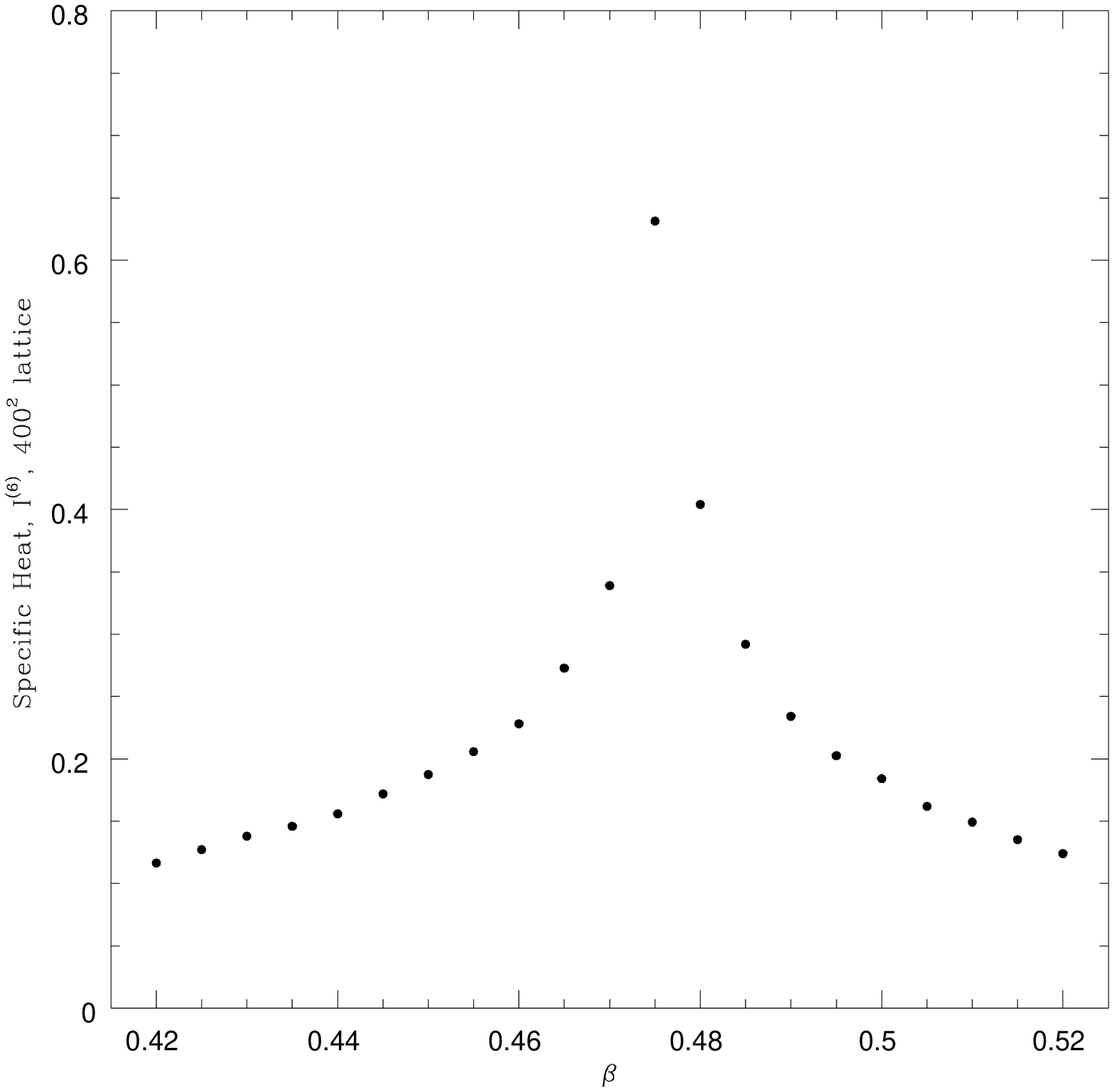}
  \caption[a]{\protect\label{F_CV6I}
    As in fig. \ref{F_CV16I}, but for the $6$ block state constrained
Ising model, $\II6$.
  }
\end{figure}

\begin{figure}
\epsfxsize=400pt\epsffile{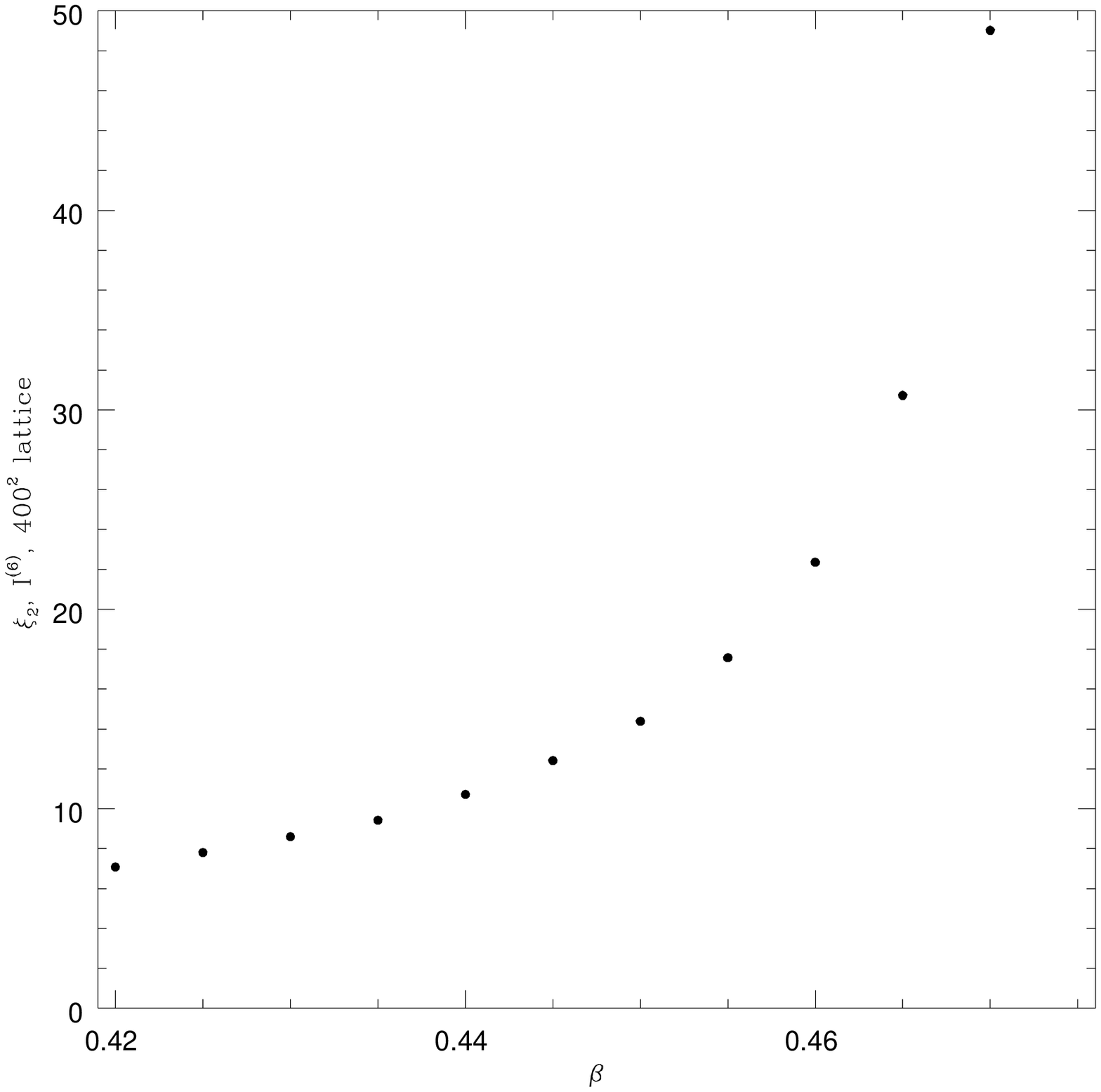}
  \caption[a]{\protect\label{F_CSI6I}
    As in fig. \ref{F_CSI16I}, but for the $6$ block state constrained
Ising model, $\II6$.
  }
\end{figure}

The second set of numerical simulations refer to model $\II6$.  We show
respectively the specific heat and the correlation length in figs.
\ref{F_CV6I} and \ref{F_CSI6I}.

These two figures strongly suggest that $T_c^{(\II6)} < T_c^{(\II{16})}$: the
critical temperature of the constrained model is smaller than the one of the
original model. At the  critical temperature of the Ising model the
constrained model does not show a critical behavior. We estimate

\be
  \beta_c^{(\II6)} = .4775 \pm .0025\ .
\ee

The correlation length of model $\II6$ is finite at the critical
temperature of the Ising model; we estimate (see \pref{corr2} for the
definition and note that it is defined in units of the original lattice):

\be
  \xi_{(2)} \equiv \xi_{(2)}(5) \simeq 11.5 \pm 0.5\ \quad,\quad \hbox{if }
  \b=\b_c^{(\II{16})}
\ee

\noindent This is in agreement with the numerical calculations related to the
$DSU$ condition, that we shall discuss in the following section. In fact we
find
that the inequality \pref{DS} of section \ref{S_INEQUA} is satisfied for
$L>L_c$, with $L_c \simeq 5 \xi$.

\begin{figure}
\epsfxsize=400pt\epsffile{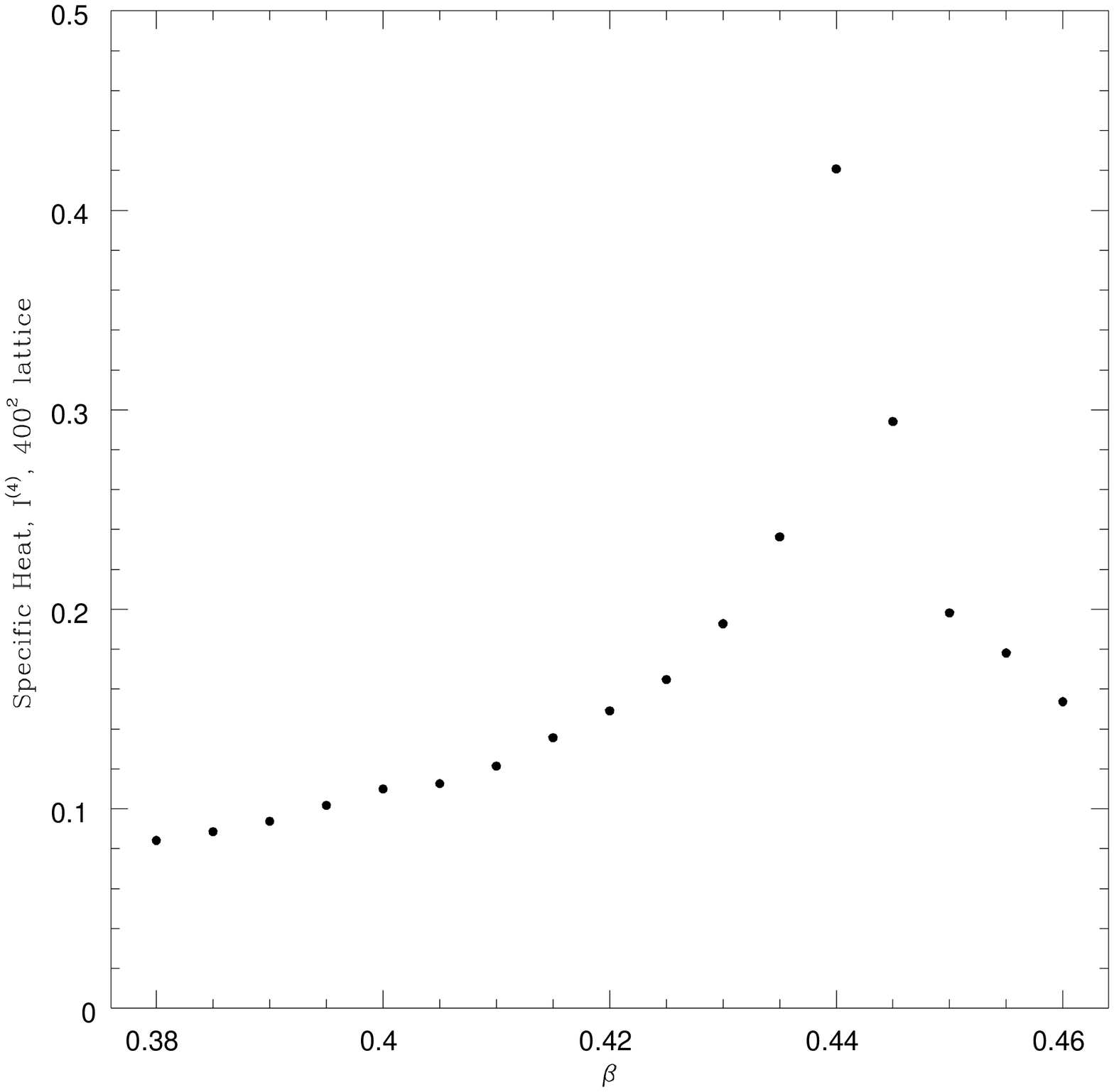}
  \caption[a]{\protect\label{F_CV4I}
    As in fig. \ref{F_CV16I}, but for the $4$ block state constrained
Ising model, $\II4$.
  }
\end{figure}

\begin{figure}
\epsfxsize=400pt\epsffile{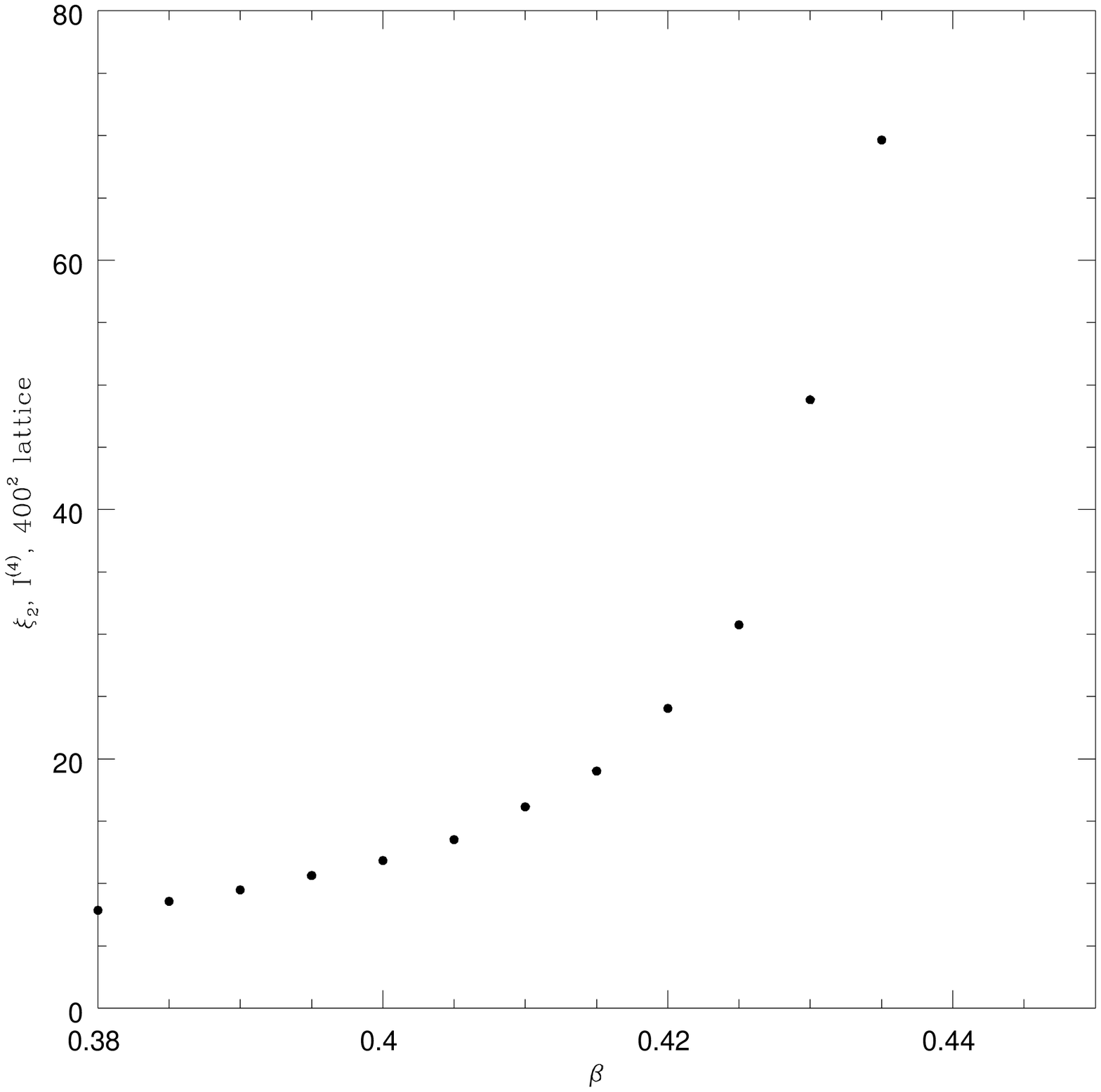}
  \caption[a]{\protect\label{F_CSI4I}
    As in fig. \ref{F_CSI16I}, but for the $4$ block state constrained
Ising model, $\II4$.
  }
\end{figure}

As we have already said, we have also simulated the model restricted to $4$
states
(by forbidding turnons), which is equivalent to two decoupled Ising models.
In figures \ref{F_CV4I} and \ref{F_CSI4I} we show the specific heat and the
correlation length; they clearly show criticality at $T_c^{(\II4)} =
T_c^{(\II{16})}$.  Note that the correlation length is larger than in model
$\II{16}$, as one expects because the effective lattice spacing of the two
independent Ising models is half the spacing in the original lattice.  The
ratio between the two correlations lengths should indeed be exactly $2$, if we
could calculate the limit of $\xi_{(2)}(t)$ as $t\to\infty$.

In order to better characterize the nature of the transition for the
constrained
model $\II6$ we will present some more data. As discussed in section
\ref{S_MOD}, at $T=0$ the $\II6$ model has $4$ ground states, with broken
translational symmetry.  We define $\rho_i$, $i=1,\ldots,4$ the projection of a
given configuration over the $i-th$ ground state (we count the number of blocks
which fit the ground state pattern, and we normalize with the total number of
blocks). We define:

\be
  \rho \equiv \frac{1}{4} (\rho_1 + \rho_2 + \rho_3 + \rho_4) \ .
\ee

\begin{figure}
\epsfxsize=400pt\epsffile{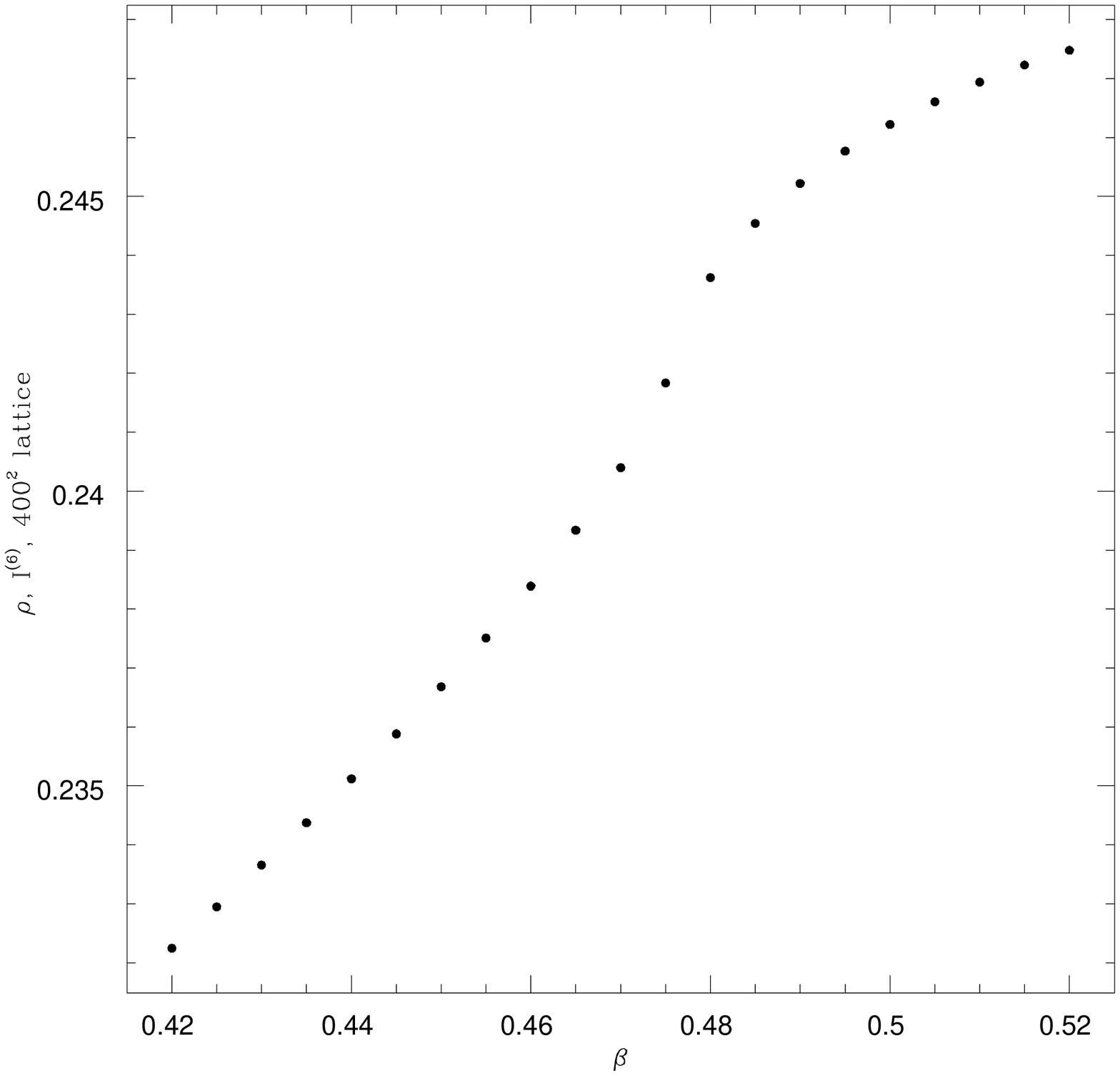}
  \caption[a]{\protect\label{F_TURN}
   $\rho$ as a function of $\beta$ for the $6$ block state model
$\II6$.
  }
\end{figure}

\noindent Let us remind again that the ground states do not include turnons.
The
turnon density is $(1-4\rho)$, and tends to zero for $\b\to\infty$. In fig.
\ref{F_TURN} we plot $\rho$ as a function of $T$ for model $\II6$. The density
of turnons at $\beta=.4407$, the critical point of the full $2D$ Ising model,
is
close to $.06$, and is determining the finite correlation length.

Spontaneous symmetry breaking is signaled by the non-vanishing, in
the infinite volume limit, of $\sigma$ defined by

\be
  \sigma \equiv \frac{1}{4}\sum_{i=1}^4(\rho_i-\rho)^2\ .
\ee

\begin{figure}
\epsfxsize=400pt\epsffile{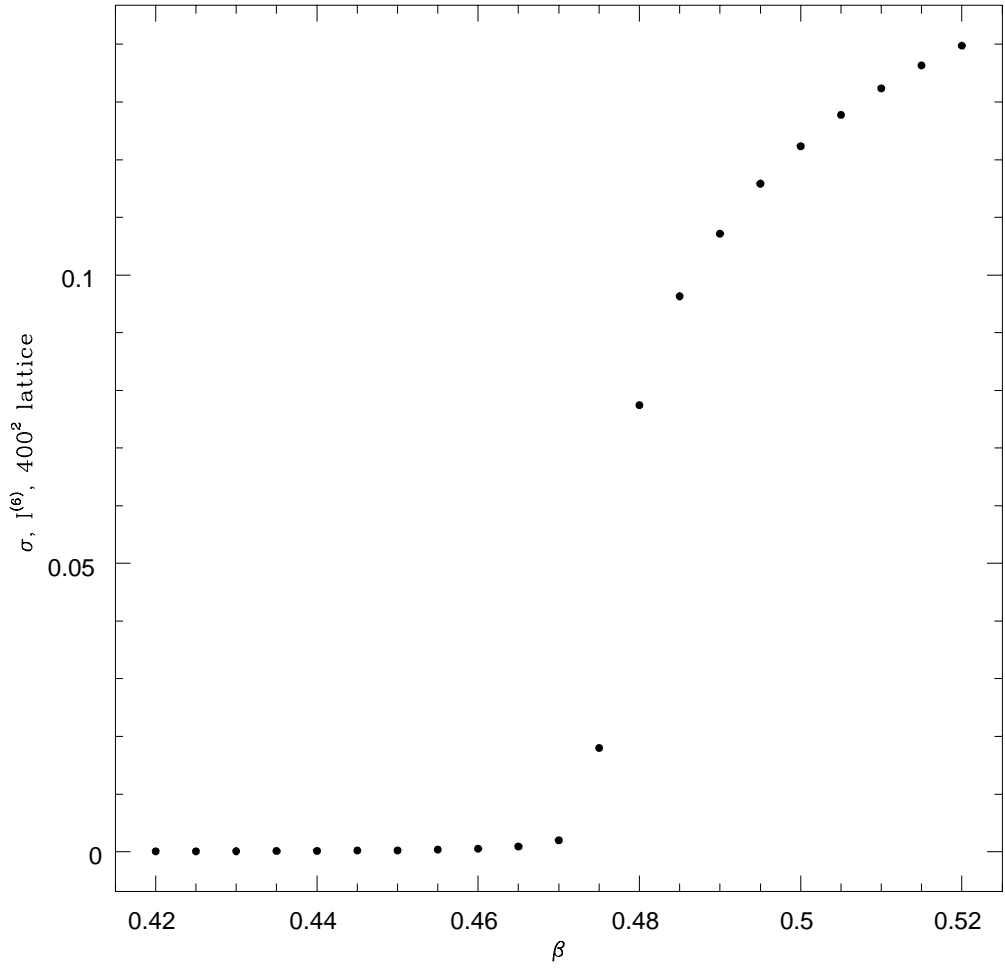}
  \caption[a]{\protect\label{F_SIG}
   $\sigma$ as a function of $\beta$ for the $6$ block state model
$\II6$.
  }
\end{figure}

\noindent In the symmetric phase $\sigma\to 0$ in the infinite volume
limit. In the broken phase the system aligns in one of the $4$ ground
states (tunneling, at finite volume, among the different layered
ground states), and $\sigma$ is non zero. We plot $\sigma$ in fig.
\ref{F_SIG}. The location of the critical point is signaled with high
precision from the drastic change in $\sigma$.

\section{\protect\label{S_BC}
Numerical Results: the Dobrushin-Shlos\-man Condition}

As discussed at the end of section \ref{S_INEQUA} we have used the inequality
\pref{DS} to obtain some reasonable insight about the validity of the
Dobrushin-Shlosman uniqueness condition, which is too difficult to verify also
numerically for volumes as large as needed in our case. However, even the task
of checking numerically the inequality \pref{DS} is indeed a difficult one,
with
respect to the calculations of the previous section. In fact in this case one
has to control the whole distribution. This is quite difficult, if
compared with the simple task of computing averages of some observable
quantities. The internal
energy, for example, is peaked close to expectation value; hence, in order to
compute its expectation value, one just needs to explore a very restricted part
of the phase space. A computation of ${\cal M}_L$ (see \pref{DS}) demands, on
the contrary, as we will see, a very large statistics.

In our simulations we have considered a sequence of cubes of side size $L$ and,
after selecting $L$ and the number $I$ of Monte Carlo sweeps (we define a Monte
Carlo sweep as the update of all lattice sites), we have chosen $N$
different random boundary condition, by assigning to the boundary sites the
value $+$ or $-$ with equal probability. We shall call $\t_k$, $k=1, \ldots, N$
these boundary conditions.

For each $\t_k$ we have selected (randomly) a boundary couple of adjacent sites
belonging to the same boundary block and we have considered the four boundary
conditions $\t_k^j$, $j=1,\ldots,4$, differing from each other only in the
chosen boundary block, by changing in all possible ways the values of the two
spins. We have done $4$ Monte Carlo runs with the four boundary conditions and,
during these runs, we have recorded the number of times each block variable has
visited each of the $6$ allowed states, by constructing

\be
  \tilde{N}_{\t_k^j}(i,S_i)\ ,
\ee

\noindent where $j$ ranges over the $4$ different values,
$i$ ranges over the $(L/2)\times (L/2)$ blocks and
$S_i$ ranges over the $6$ allowed block configurations.
$\tilde{N}$ is normalized in such a way that

\be
  \sum_{S_i} \tilde{N}_{\t_k^j}(i,S_i) = 1\ .
\ee

\noindent The quantity $\D_{L/2}$ of \pref{DS} was then calculated as

\be
  \D_{L/2} \equiv \max_{k=1,\ldots,N}\  \max_{j\not=l; j,l=1,\dots,4}
  \{ \frac{1}{2} \sum_{i,S_i}
  |\tilde{N}_{\t_k^j}(i,S_i) - \tilde{N}_{\t_k^l}(i,S_i) | \}\ ,
\ee

In our simulations we have used $L=8$, $16$, $32$, $64$, $N$ of order
$100$ and values of $I$ ranging from $4000$ to $4 \ 10^6$ (for the
largest value of $L$).

\begin{figure}
\epsfxsize=400pt\epsffile{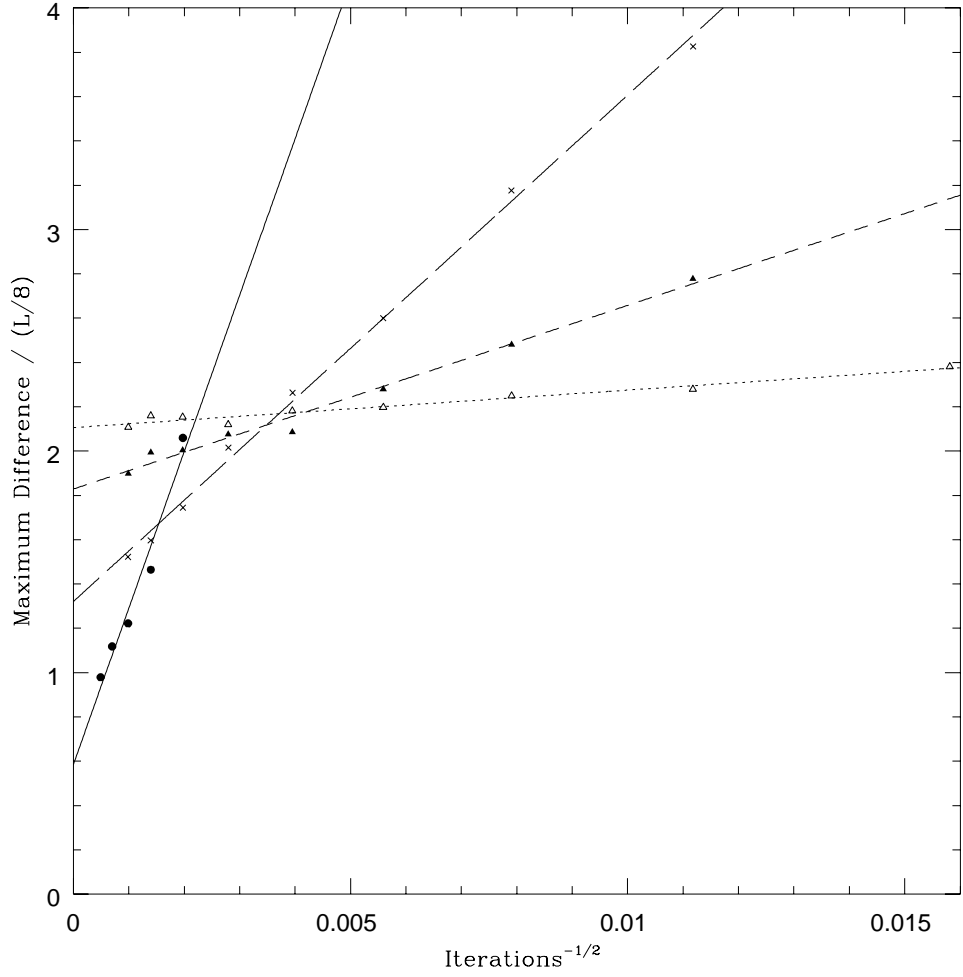}
  \caption[a]{\protect\label{F_SUM}
    ${\cal M}_L$
as a function of the inverse square root of the number of
sweeps, for different lattice sizes. Straight lines are best linear
fits. Empty triangles for $L=8$, filled triangles for $L=16$, crosses
for $L=32$ and filled dots for $L=64$.
  }
\end{figure}

In fig. \ref{F_SUM} we plot the quantity ${\cal M}_L$ of \pref{DS} as a
function
of the inverse square root of the run length, for different lattice sizes and
number of iterations. The straight lines are our best linear fit, which turns
out to be the right ansatz for the observed behavior.

The inequality \pref{DS} is satisfied only on the $64^2$ lattice, and it is
violated on smaller lattices. Our best fits give

\bea
  {\cal M}_{8}  & = & 2.10 + \frac{ 17}{\sqrt{I}}\ , \\
  {\cal M}_{16} & = & 1.83 + \frac{ 83}{\sqrt{I}}\ , \\
  {\cal M}_{32} & = & 1.32 + \frac{229}{\sqrt{I}}\ , \nonumber\\
  {\cal M}_{64} & = & 0.58 + \frac{706}{\sqrt{I}}\ .\nonumber
\eea

\begin{figure}
\epsfxsize=400pt\epsffile{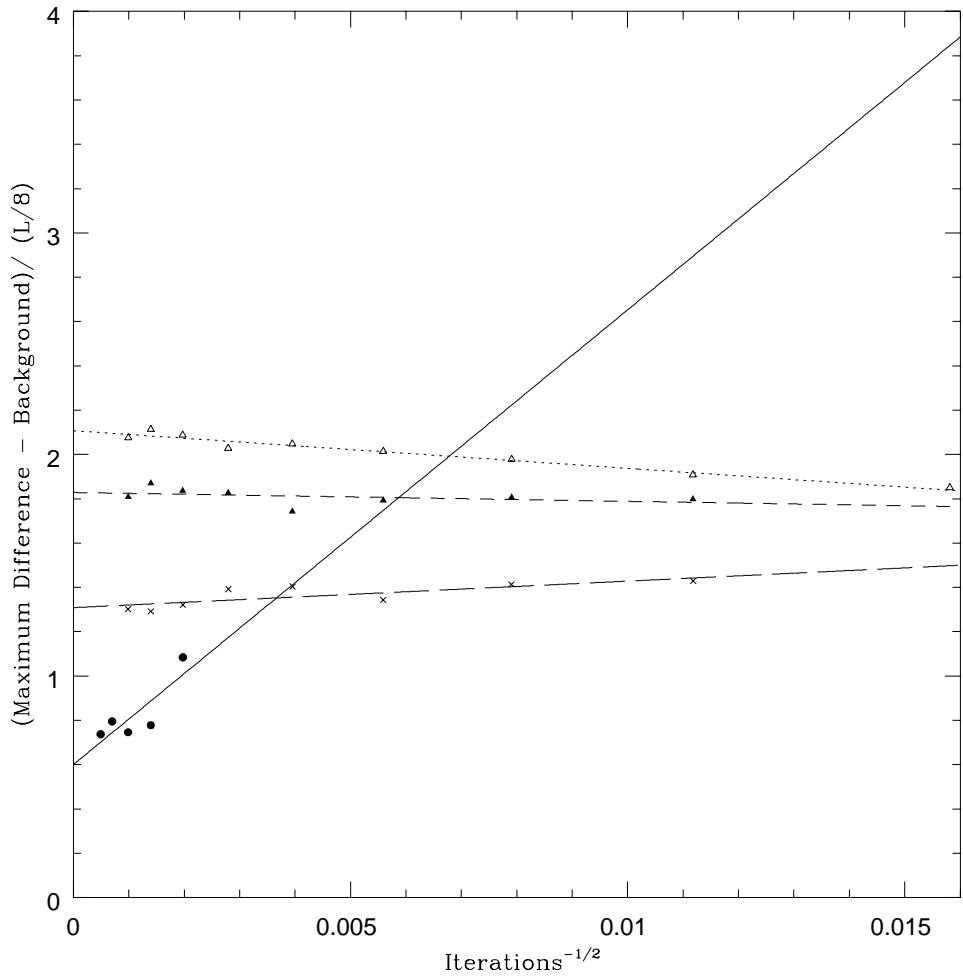}
  \caption[a]{\protect\label{F_SUMS}
    As in fig. \ref{F_SUM}, but for $\tilde{\cal M}_L$.
  }
\end{figure}

\noindent There is indeed a big contribution due to the fact that we
are adding a finite number of positive random numbers. Only in the
limit of large number of iterations, $I\to \infty$, this contribution
goes to zero. In order to minimize this effect we have also run
simulations with the same boundary condition $\t_k$, for each $k$.
The average value of the difference

$$
  \{ \frac{1}{2} \sum_{i,S_i}
  |\tilde{N}_{\t_k}(i,S_i) - \tilde{N'}_{\t_k}(i,S_i) | \}
$$

\noindent obtained in these conditions has been subtracted from ${\cal M}_L$ in
order to define $\tilde{\cal M}_L$.  The contribution we have subtracted from
${\cal M}_L$ goes to zero in the limit $I\to \infty$, making $\tilde{\cal
M}_L$ a good estimator for the inequality \pref{DS}.  We plot $\tilde{\cal
M}_L$ in fig.~\ref{F_SUMS}.  Our best fits for $\tilde{\cal M}_L$ give

\bea
  \tilde{\cal M}_{8}  & = & 2.11 - \frac{ 17}{\sqrt{I}}\ , \\
  \tilde{\cal M}_{16} & = & 1.83 - \frac{4.0}{\sqrt{I}}\ , \\
  \tilde{\cal M}_{32} & = & 1.31 + \frac{12.0}{\sqrt{I}}\ , \nonumber\\
  \tilde{\cal M}_{64} & = & 0.60 + \frac{205}{\sqrt{I}}\ .\nonumber
\eea

\noindent The constant term coincides, as it should, with the one we get for
$\cal M$, with very good precision. On the contrary slopes are
smaller, indicating that $\tilde{\cal M}$ is, as expected, a better
estimator than $\cal M$ when using a finite number of iterations.

\section{\protect\label{S_CON} Conclusions}

 From the numerical results on specific heat and correlation length, which are
based on ``traditional'' numerical methods to detect a (second order) phase
transition, we can be reasonably sure that, indeed, the critical temperature of
$\II6$ model is strictly less than $ T_c^{Ising}$. The remark on the
isomorphism
between $\II4$ and the full Ising model $\II{16}$ shows that one cannot, a
priori, expect any monotonicity property of critical temperatures in terms of
imposed constraints. Our numerical results indicate that, by imposing the
constraint $m_i =0$, $\forall \; B_i$, one  {\it decreases} the critical
temperature whereas, by enhancing again the constraint via the further
elimination of the turnons, one {\it increases} the critical temperature since
it goes back to $T_c^{Ising}$.

To theoretically analyze this apparently strange behavior it
seems useful to use some generalized form of Fortuin-Kasteleyn
representation of the Ising model sufficiently ``elastic'' to
include, in the same set-up of random-cluster models, the three
models $\II{16}$, $\II6$, $\II4$ that we have considered (\cite{G}).

On the other hand to rigorously implement the Cassandro-Gallavotti
program one needs a {\it strong} notion of absence of phase
transition, namely to verify some strong mixing condition SM.

We say that a Gibbs measure $\mu_{\Lambda}^{\tau}$ in $\Lambda$
with $\tau$ boundary conditions outside $\Lambda$ satisfies a
{\it strong mixing condition } $SM(\Lambda,C,\gamma)$ if the
influence at $x\; \in \; \Lambda$ of a {\it local} change in
 $y\; \in \; \Lambda^c$ of the value of the conditioning spin
configuration $\tau$ decays as

$$
  C e^{- \gamma |x-y|}
$$

\noindent Of course  $SM(\Lambda,C,\gamma)$ implies
$WM(\Lambda,C,\gamma)$ (see section \ref{S_INEQUA}); moreover
$SM$ $(\Lambda,$ $C,$ $\gamma)$  is interesting
when it is valid for a class of volumes $\Lambda$ invading ${\bf
Z^d}$ with $C$ and $\gamma$ independent of $\Lambda$.

As we have said in section \ref{S_GD}, in \cite{MOS} it is proven that, in two
dimensions, $WM$ implies $SM$ at least for sufficiently regular regions
$\Lambda$. $SM$ would be certainly largely sufficient to compute, via the
Cassandro-Gallavotti method, the renormalized potentials whereas, in general,
$WM$ alone will not.

In section \ref{S_BC} we have analyzed a sort of lower bound (involving
variation
distance) for the quantity appearing in the Dobrushin-Shlosman uniqueness
condition (implying $WM$ and so, since we are in $2D$, $SM$).  This is only an
(almost) necessary condition to be verified in order to satisfy the true
condition
$DSU$.  It only gives an indication in the sense of the possibility to verify
$DSU$ (which involves Vasserstein distance).

It is clear that one needs to consider squared regions containing
about $30$ $\times$ $30$ blocks.  This rules out, at least with the present
time
computers, the possibility of a computer-assisted proof.  Then it would be
important to be able to find a ``Montecarlo'' method to ``measure'' the
quantity (Vasserstein distance) appearing in $DSU$.

It would also be interesting, in general, to find an algorithm, easily
implementable in a computer, to evaluate the Vasserstein distance
between two Gibbs measures with different boundary conditions.
This will be the object of further investigations.

\section*{Acknowledgments}

We are indebted with Giovanni Gallavotti and Marzio Cassandro for very
stimulating and illuminating discussions and for their active participation to
the early stage of the present work.  We acknowledge an inspiring discussion
with Giorgio Parisi about the issues raised in this note. We want to thank
Charles
Pfister and Aernout van Enter for clarifying to us very important points
related to our work.


\newpage

\end{document}